# The use of segmented cathode of a drift tube for designing a track detector with a high rate capability

N.A.Kuchinskiy, V.A.Baranov, V.N.Duginov, F.E.Zyazyulya*, A.S.Korenchenko, A.O.Kolesnikov, N.P.Kravchuk, S.A.Movchan, A.I.Rudenko, V.S.Smirnov, N.V.Khomutov, V.A.Chekhovsky*

Joint Institute for Nuclear Research

Russia, 141980 Dubna, Moscow Region, Joliot-Curie, 6

* National Scientific and Educational Center of Particle and High Energy Physics, Belarusian State University, Minsk

## Abstract

*Detector rate capability is one of the main parameters for designing a new detector for high energy physics due to permanent rise of the beam luminosity of modern accelerators. One of the widely used detectors for particle track reconstruction is a straw detector based on drift tubes. The rate capability of such detectors is limited by the parameters of readout electronics. The traditional method of increasing detector rate capability is increasing their granularity (a number of readout channels) by reducing the straw diameter and/or by dividing the straw anode wire into two parts (for decreasing the rate per readout channel). A new method of designing straw detectors with a high rate capability is presented and tested. The method is based on dividing the straw cathode into parts and independent readout of each part.*

A new method to create straw detectors with a high rate capability is presented and tested. The method is based on the division of the straw cathode into parts and independent readout of each part. This method allows setting up inside a straw's volume several "elementary detectors" with independent readout, which provides the high rate capability of the entire straw. Readout from "elementary detectors" is from both straw ends. Besides, no additional component is introduced into the straw's volume. Therefore, the straw detector "transparency" and registration efficiency are the same. There is no need to reduce the straw diameter. The total number of readout channels remains the same as for traditional solutions. The proposed method is technologically simple due to ultrasonic welding of plastic tape along the straw axis.

Detectors based on thin-walled drift tubes (straws) with the diameter of 4 to 10 mm are widely used today as coordinate detectors. For example, straw detectors are used in such experiments as SDC, ATLAS, COMPASS [1-3]. The advantages of such detectors are: high spatial resolution of about 100 µm (based on drift time measurement), track reconstruction efficiency is quite near 100%, high rate capability of about 500 kHz per readout channel, simple detector design and, consequently, its low cost. In addition the cylindrical geometry of the straw allows to have really good mechanical characteristics with low amount of material. Each straw is self-sufficient: a failure in the operation of one of the straws doesn't have any influence on operation of the rest ones.

Due to these advantages the number of straws used in the modern experiments permanently increase. For example, the ATLAS TRT detector has about 300 000 straws with a diameter of 4 mm [4]. A straw can reach up to 4 meters in length. [5].

At present with the rise of beam luminosity of modern accelerators detector rate capability becomes the most important parameter. High rate capability can be achieved by development of new high rate detectors or by reduction of detector sensitive area and corresponding increase in granularity. In respect to straws this requirement is usually solved by reducing the straw diameter. Another solution consists in electrical division of the anode wire into sections with independent readout of each section [6]. This solution was suggested and implemented in the ATLAS TRT detector [4], where the straw anode wire was divided into two electrically independent pieces by using glass insulator with a length of 7 mm and a diameter of 0.25 mm. This allows having two independent detectors in the same straw volume with readout at both ends of the straw. Rate load on readout channel was decreased by factor of 2.

The idea of the anode wire dividing into many electrically independent pieces with individual readouts and high voltage power applying was suggested and implemented [7,8]. The length of each anode piece can reach few centimeters [9]. The signals from separate pieces are transferred through the straw wall with the help of specially designed spacers. The same spacers were used to apply high voltage to the anode segments. We shall note that the solution degrades such important features of the straw detector as "transparency" and registration efficiency due to big amount of material (large spacers) introduced in the straw.

We propose to solve the problem of increasing the straw rate capability by dividing the cathode surface into electrically isolated segments. The signal from the anode avalanche induces the signal on one or two cathode segments depending on the place of avalanche with respect to the line of electric isolation of the segments. Thus, several relatively independent detectors are used in the volume of one straw. The information from the cathode segments can be used for localization of radial track position and localization of track position along the straw. Radial track position can be measured using drift time of the first electron from primary particle ionization. The localization of track position along the straw can be done by simple identification of the triggered segment or using charge division method for two neighboring segments. In our study we used straws with a diameter of 10 mm and a length of 300 mm. Straws were manufactured from one side metallized 36 μm thick mylar. The metal coating is two-layer: copper and gold thickness are 0.05 and 0.02 μm, respectively. The sheet resistance is 40 ohm/square. The straw cathode was divided into two electrically isolated segments in the middle. The signals were read out from both straw ends. Segments were preliminarily cut out on the tape coating and the straw was manufactured by the method of ultrasonic welding. The straw manufacture technology is presented in reference [10].

Straw tubes manufactured by ultrasonic welding have thinner walls compared to ones made by traditional method (coiling tape on a rod). Such straws can operate at overpressure up to 8 bar and under vacuum.

The shape of the cathode segments is shown in Fig.1. The straw diameter is about 10 mm and the length – about 300 mm. The straw length was limited by available test-bench mechanics. The anode wire diameter is 30 μm. Gas mixture of Ar/$CO_2$ (70:30) was used at overpressure of 10 mbar. High voltage was about 3200 V.

The measurement block-diagram is presented in Fig.2. $^{90}$Sr radioactive source was used. Beta particles penetrated a straw and were detected in a trigger scintillation counter. The straw was moved by a precision manual positioner with respect to the radioactive source and the counter.

The scintillation counter is based on BICRON fiber with a 2 mm square cross section and a length of 5 mm. A CPTA 149-35 SiPM (4.41 $mm^2$ sensitive area, 1764 pixels [11]) is used for light detection. A lead collimator with a 1 mm × 5 mm slit was used. The slit was perpendicular to the straw axis.

Cathode and anode signals were amplified by KATOD-1 [12] and Ampl. 8.3 [13] chips. Signal shapes were digitized by a 12bit 250 MS/s CAEN V1720 digitizer [14] connected to a PC via VME.

Straws with 2 and 6 segments (see Fig.1) were tested. Segment signals were delivered to the straw ends by strip lines on the straw cathode.

Typical signal shapes from two cathode segments are shown in Fig. 3 for different positions of the radioactive source: at the point of the electrical isolation between the segments and at a distance of 40 mm from that point. Efficiency versus the z-coordinate along the straw axis is presented in Fig.4. The segment was considered effective if the amplitude of its signal was bigger than 25 mV.

The efficiency for the straw with 6 segments is shown in Fig.5. A 100 mm long part of the straw (z = 75-175 mm) with the cathode segments 2B-5B was scanned. In this case the efficiency threshold value of 50 mV was set according to measurements of the segment efficiency versus threshold value (see Fig.6) and the segment signal amplitude distribution (see Fig.7).

Studying the possibility of track radial coordinate determination by drift time measurement with the use of signals from the cathode segments is a challenging topic. Drift time distributions for the anode and cathode segment signals with respect to the scintillation trigger were measured. The results are presented in Fig.8. The similarity of the shapes of the distributions provides a possibility for a precise measurement of the track coordinates using the drift time obtained from the arrival time of cathode signals.

So, it is possible to produce straws with cathode segments having independent readouts. In case of nonuniform rate along the straw it is possible to adjust the rate on each segment by optimization of the straw segmentation.

As it was mentioned above the segment signals were delivered to the straw ends with the strip lines on the straw cathode. The strip width was 1 mm. The total number of such strips is limited. The problem of segment signals delivery to the straw ends can be solved by making holes in the straw cathode. For example, conductive glue can be used for electrical connection of the cathode segments with the pads on the straw outer surface. The use of cathode segments and segment-pad contacts give broad options for various ways of readout from the cathode segment.

The proposed method of increasing straw rate capability is based on using cathode segments. The advantages of such a solution are: high transparency for particles, capability to obtain two coordinates at once and efficient technology of straw manufacture. At that, as it was shown, the small area of the straw welding seam without metal coating doesn't influence the straw operation and precision of track coordinate measurement by drift time.

**Summary**

The method of increasing straw rate capability based on using electrically isolated cathode segments with individual readout for each of them was proposed and tested. This approach is characterized by simple and efficient manufacturing technology while retaining all the inherent advantages of straw detector. Straw of such type can be manufactured by ultrasonic welding from preliminarily prepared tape with segments and strip lines for segment readout.

The investigation was supported by the Russian Foundation for Basic Research grant 13-02-00745 and the Belarusian Republican Foundation for Fundamental Research grant BRFFR-JINR F12D-10.

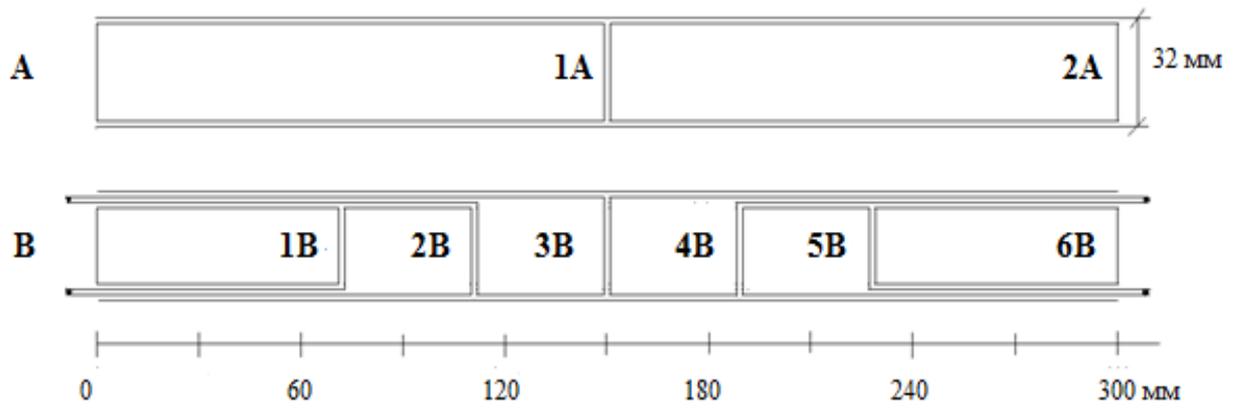

Fig.1: Straw cathode drawings with 2 (A) and 6 (B) segments.

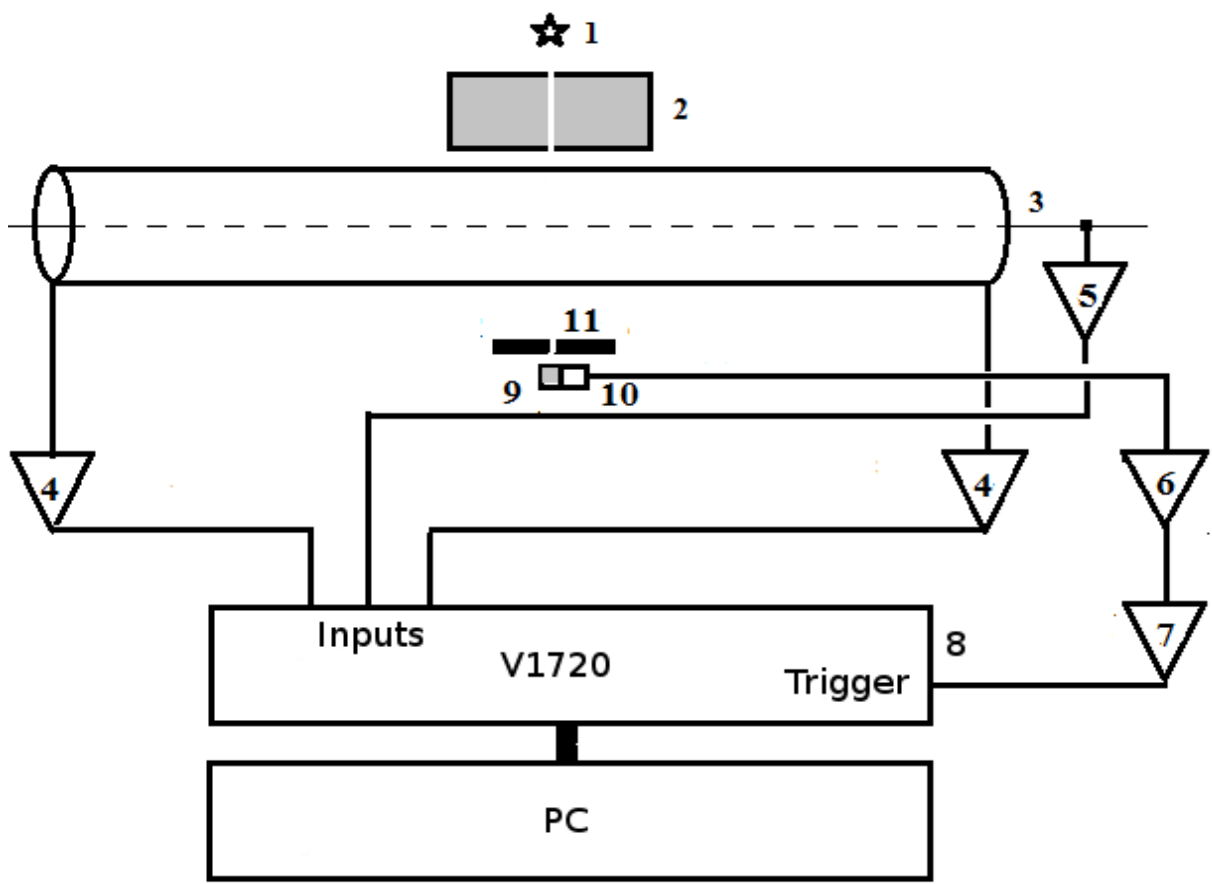

Fig.2: Block-diagram of the set-up for straw parameters study. 1 – $^{90}$Sr, 2 – slit collimator, 3 – anode wire ( Ø 30 μm), 4 – cathode preamp., 5 – anode preamp., 6 – SiPM preamp., 7 – discriminator, 8 – digitizer, 9 – scintillation counter 2×2×5 mm, 10 – SiPM, 11 – slit collimator (1×5 mm).

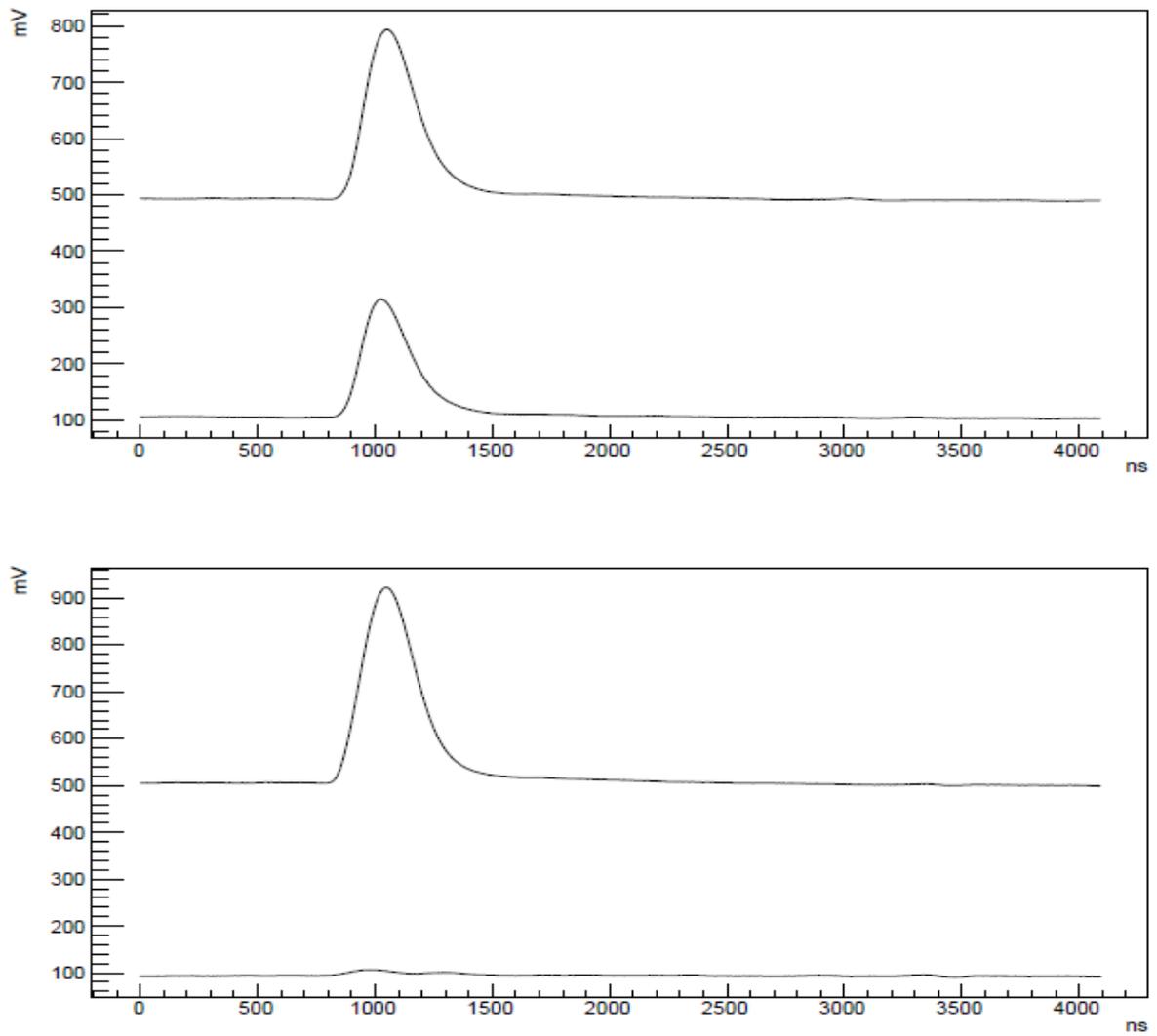

Fig.3: Signal shapes from 2 cathode segments. Top – $^{90}$Sr radioactive source is between 2 cathode segments, bottom – $^{90}$Sr source is near one of the segments.

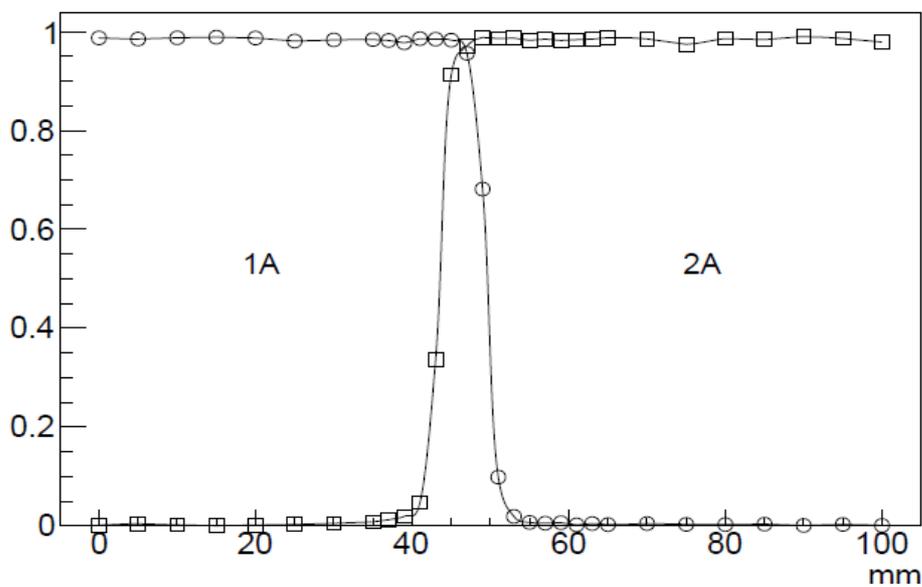

Fig.4: Efficiency versus the Z-coordinate along the straw axis for collimated $^{90}$Sr source. Discriminator threshold = 30 mV. HV = 3200 V.

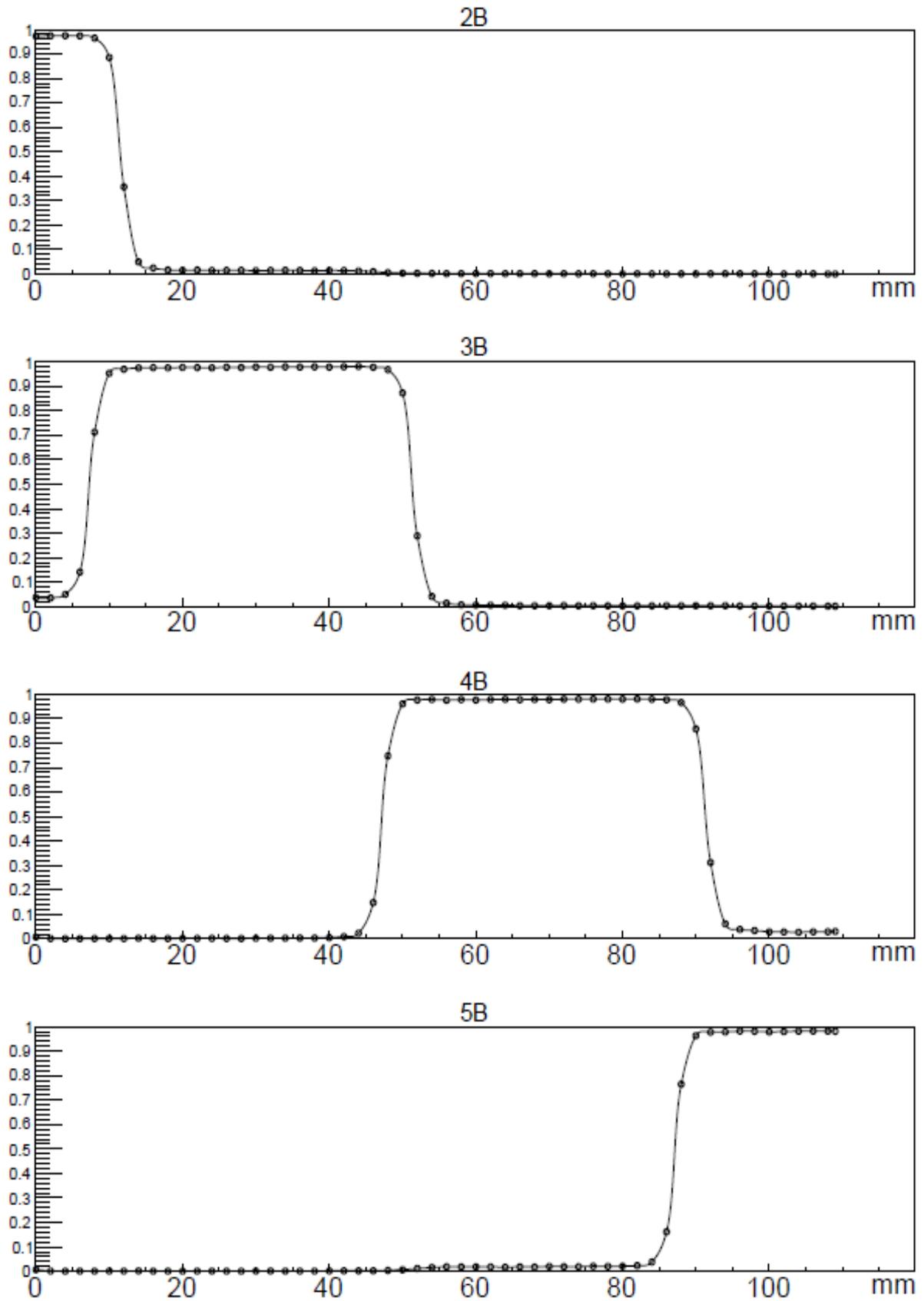

Fig.5: Segment efficiency versus the Z-coordinate along the straw axis (Z = 75-175 mm) for the collimated $^{90}$Sr source. The data are shown for the 4 middle cathode segments of the 6-segment straw. Discriminator threshold = 50 mV. HV = 3200 V.

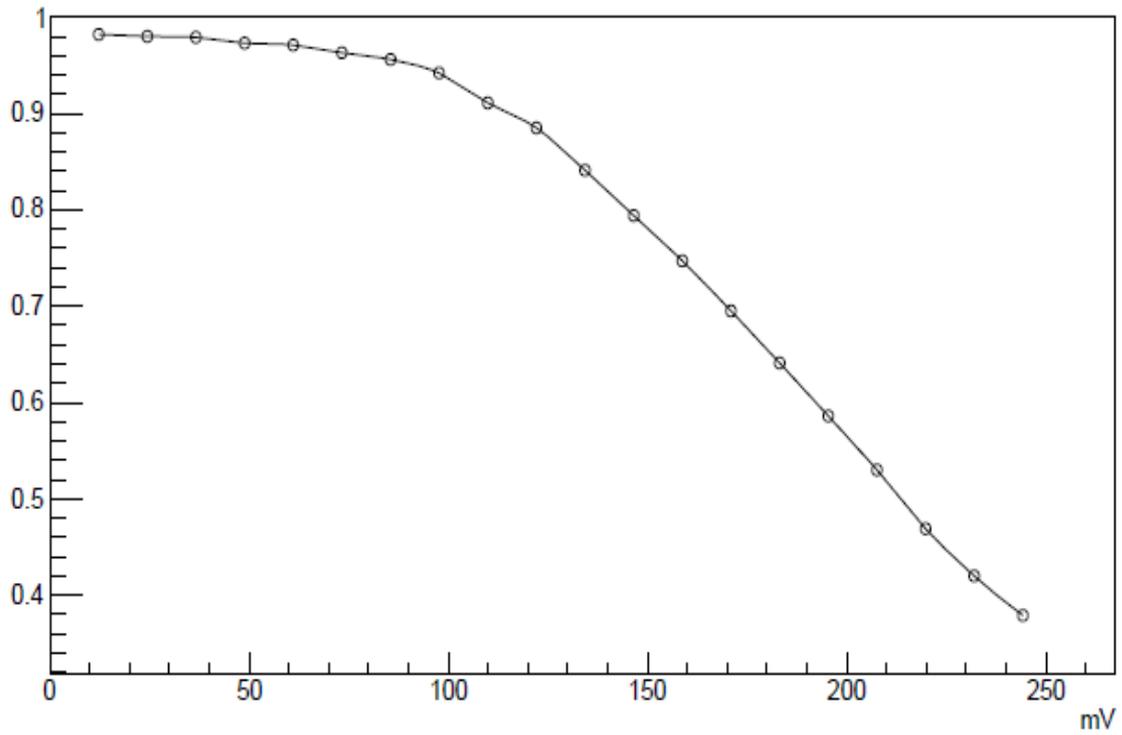

Fig.6: Segment efficiency versus the discriminator threshold at HV = 3200 V.

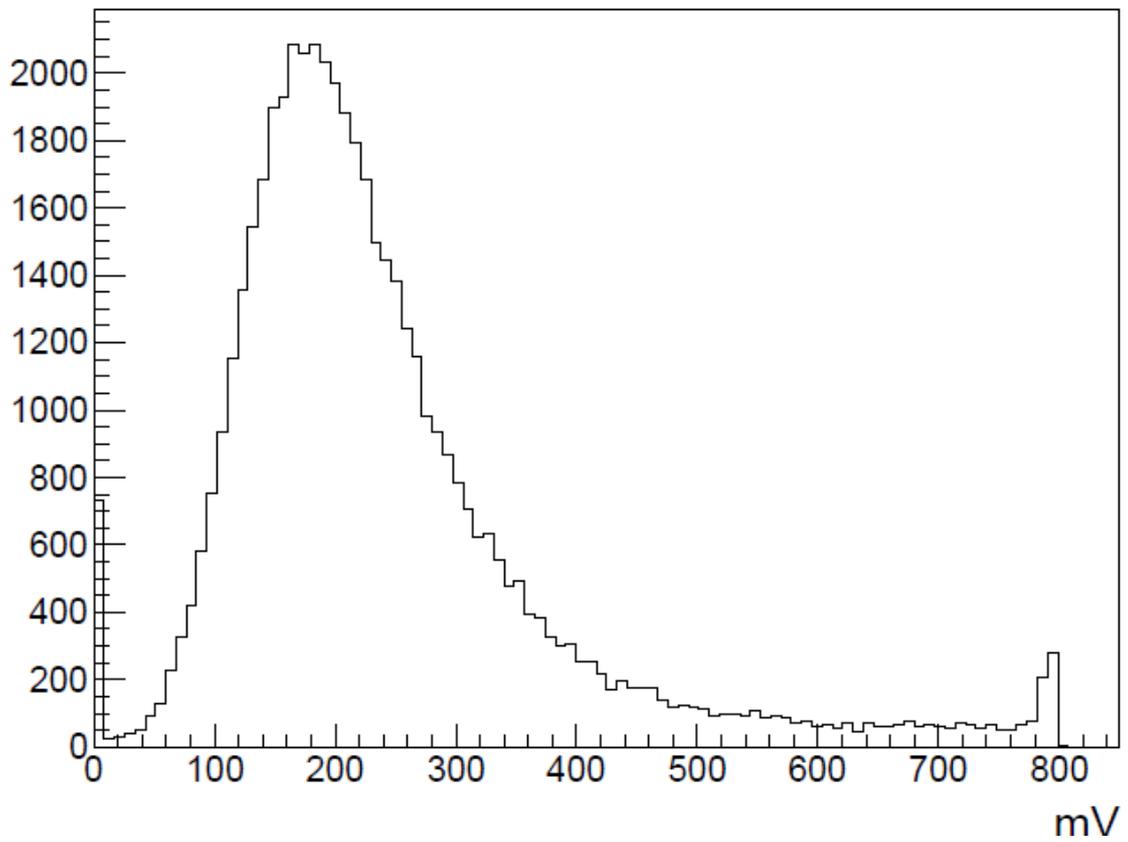

Fig.7: Segment signal amplitude distribution at HV = 3200 V.

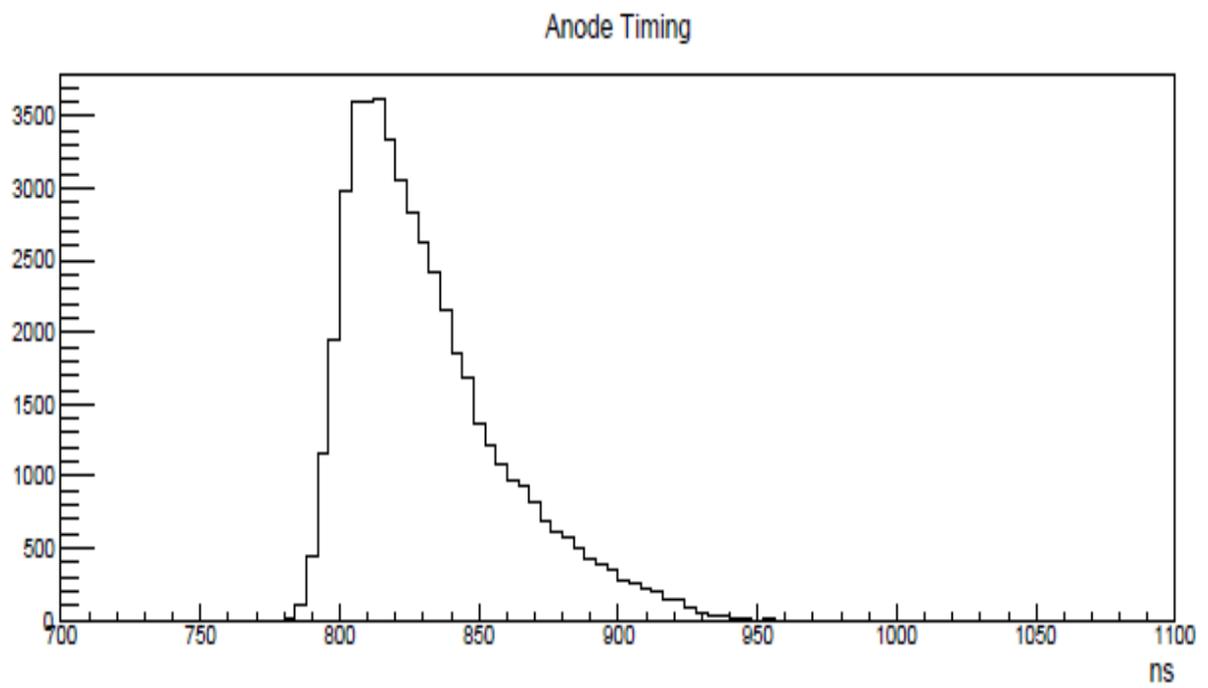

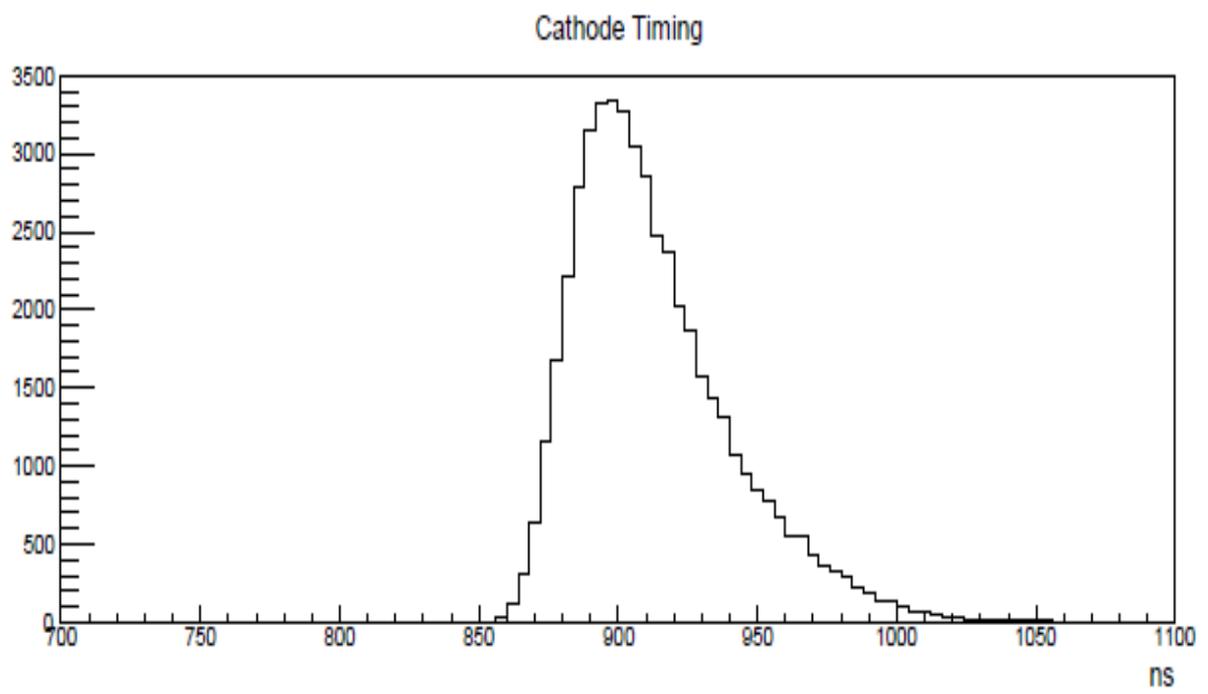

Fig.8: Drift time distributions for anode wire (top) and cathode segment (bottom) signals.